\documentstyle[prl,aps]{revtex} 
\input{psfig.sty} 

\tighten

\begin{document}
\draft

\twocolumn[
\hsize\textwidth\columnwidth\hsize\csname @twocolumnfalse\endcsname

\title{Dynamics of Fluxon Lattice in Two Coupled Josephson Junctions}

\author{ A. F. Volkov$^{\dagger*}$, V.A. Glen$^{*}$ }

\address{$^{\dagger}$ Institute of Radioengineering and Electronics of the
Russian Academy of Sciences, Moscow}

\address{$^{*}$ School of Physics and Chemistry, Lancaster University,
Lancaster LA1 4YB, U.K.}

\date{\today}	
\maketitle

\begin{abstract}
We study theoretically the dynamics of a fluxon lattice (FL) in two coupled
Josephson junctions. We show that when the velocity of the moving FL exceeds
certain values $(V_{a,b})$, sharp resonances arise in the system which are
related to the excitation of the optical and acoustic collective modes. In the
interval$(V_a,V_b)$ a reconstruction of the FL occurs. We also  establish that
one can excite localized nonlinear distortions (dislocations) which may
propagate through the FL and carry an arbitrary magnetic flux.
\end{abstract}

%%\pacs{Pacs numbers: 72.10Bg, 73.40Gk, 74.50.+r}
] \narrowtext

In recent years many papers have addressed the dynamics of the fluxon lattice in
layered superconductors and in particular, in high $T_c$ superconductors. In the
absence of the magnetic field, the spectrum of collective oscillations in the
long  wave limit has a threshold frequency which coincides with the Josephson
frequency $\omega_J$ \cite{r1} - \cite{r7}. If a magnetic field is applied, the
vortex lattice is formed in a superconductor, and the spectrum of collective
modes is changed. In particular, an acoustic-like mode related to the vortex
lattice  oscillations arises. The spectrum of the fluxon lattice (FL) arising in
a parallel magnetic field was calculated in Refs.\cite{r6} - \cite{r8}.

The collective modes  may be excited by an external ac field and also by a dc
current $(j_{dc})$ across the layers. In the latter case, the collective modes
are excited if the velocity of the moving FL coincides with the phase velocity
of the collective modes. This effect has been studied in detail for the case of
a long Josephson tunnel junction \cite{r9} and was studied in a  theoretical
paper \cite{r10} recently for the case of layered superconductors.

In the present paper we consider two coupled, long Josephson junctions and study
the excitations of small amplitude collective modes as well as the excitations of
nonlinear perturbations (dislocations) of the FL in such a system by a dc
current. Comparative simplicity of equations governing the FL dynamics in this
system, allows one to analyse effects arising in this system in detail and to
understand the behaviour of more complicated structures, like layered
superconductors. It will be shown in particular that nonlinear excitations
(dislocations) may arise in the system by the dc current and that these
exciations can carry an arbitrary magnetic flux (larger or smaller than the
magnetic flux quantum $\Phi_0$). We note that the system under consideration was
analysed in the absence of a magnetic field in Refs. \cite{r11} - \cite{r13}.

{\bf I Model and Basic Equations}

Let us consider the system shown in fig.\ref{f1}. We assume that different currents may
be passed through the junctions 1 and 2, i.e. a current through the middle
superconductor can be driven independently from currents through the outer
superconducting electrodes. For simplicity we assume that the junctions are
identical, i.e. they have equal critical currents etc. Equations describing the
dynamics of the coupled Josephson junctions were obtained in a number of works
\cite{r4} - \cite{r8}, \cite{r11} - \cite{r13}. The magnitudes of the magnetic
field in junctions 1 and 2 are related to the phase difference $\varphi_{1,2}$
through the expression

\begin{equation}
H_{1,2} = (1/2) \partial_x \left[ \varphi_{1,2} +
\gamma \varphi_{2,1} \right]
\label{1}
\end{equation}

Here $H_{1,2}$ are the dimensionless magnitudes of the magnetic field in the
junctions 1 and 2. We choose the quantity $H_0 = \Phi_0 / 2 \pi \lambda^2$ as a
unit of measurement of the magnetic field and the London penetration depth as a
length unit. $\Phi_0$ is the magnetic flux quantum, $\varphi_{1,2}$ is the phase
difference in junctions 1 and 2 and $\gamma$ = exp[-2d], where 2d is the
thickness of the middle electrode (in units of $\lambda$). We assume that the
characteristic scale of spatial variations of $\varphi_{1,2}$ is much greater
than unity, (i.e. than $\lambda$) and also that the thickness of the outer
superconducting layers are greater than $\lambda$.

The current through the junctions can be expressed through the corresponding 
component of ($\nabla \times H$) and be related to the quasiparticle and
Josephson currents. We obtain

\begin{equation}
2 l^2_J \partial_x H_{1,2} = \left( \partial^2_{tt} + \alpha \partial_t \right)
\varphi_{1,2} + sin (\varphi_{1,2}) - \eta_{1,2}
\label{2}
\end{equation}

Here $l_J = (c H_0 / 8 \pi j_c\lambda)^{1/2}$ is the dimensionless Josephson
penetration length, $j_c$ is  the critical Josephson current density, $\alpha =
\hbar / 2 e \rho_{qp} j_c t_0$ is the damping constant, $\rho_{qp}$ is the
junction resistivity due to quasiparticle tunneling. Time is measured in units
$t_0 =\sqrt{\hbar C / 2e j_c}$, where C is the junction capacitance (per unit
area). The constants $\eta_{1,2}$ are dimensionless currents (in units of $j_c$)
through the junctions 1 and 2. It can easily be shown that the magnetic flux in
the system equals

\begin{equation}
\Phi = \int^L_0 dx \partial_x \left( \varphi_1 + \varphi_2 \right)
\label{3}
\end{equation}

Substituting for $H_{1,2}$ in (\ref{2}) from (\ref{1}), we obtain a set of two
coupled equations for $\varphi_{1,2}$

\begin{equation}
\l^2_J \partial^2_{xx} \left[ \varphi_{1,2} + \gamma \varphi_{2,1} \right]
= \left( \partial_{tt} + \alpha \partial_t \right)  \varphi_{1,2}
+ sin (\varphi_{1,2}) - \eta_{1,2}
\label{4}
\end{equation}

We introduce the new functions $\varphi_\pm = (1/2) (\varphi_1 \pm \varphi_2)$.
Summing and subtracting eqs.(\ref{4}), we obtain the new equations for
$\varphi_{\pm}$.

\begin{equation}
\l^2_\pm \partial^2_{xx} \varphi_\pm  =  (\partial_{tt} + \alpha\partial_t )
\varphi_\pm  + sin(\varphi_\pm) cos(\varphi_\mp) - \eta_\pm
\label{5}
\end{equation}

Here $l^2_\pm = l^2_J (1 \pm \gamma)$ and $\eta_\pm = (\eta_1 \pm \eta_2)/2$.
Eqs.(\ref{5}) describe the dynamics of two coupled Josephson junctions. We use
them for studying the FL. Let us assume that a magnetic field parallel to the
planes of the Josephson junctions is applied and a dense FL arises in  the
junctions 1 and 2. In the stationary state (and sufficiently high magnetic
fields) the solution for $\varphi_{1,2}$ can be easily found from eqs.(\ref{4}):
$\varphi^{(s)}_{1,2}$ = ${\cal H}x \pm (\pi/2)$ + $\psi^{(s)}_{1,2}$ Where
$\psi^{(s)}_{1,2} \approx \mp (l_- {\cal H})^{-2} cos({\cal H}x)$. This solution
is valid provided

\begin{equation}
\l^{-1}_-  \ll  {\cal H } \ll  2\pi
\label{6}
\end{equation}

Then the right hand side of this condition means that eq.(\ref{1}) is valid,
i.e. the characteristic scale of $\varphi_{1,2}$ variation along the x-axis is
greater than the London penetration depth. The left hand side of this condition
means that the FL is dense and spatial oscillations of $\varphi_{1,2}$ are
small. The field in the junctions $\cal H$ is related to the external field by
${\cal H} = 2H_e/(1+\gamma)$. The expression for $\varphi^{(s)}_{1,2}$ given
above describes two fluxon chains each of which  is shifted by a half period
with respect to each other.

{\bf II Dynamics of the Fluxon Lattice}

Let us now consider solutions describing the motion of the FL driven by the dc
currents $\eta_{1,2}$ (the currents $\eta_{1,2}$ may differ from each other). We
seek the solution of eqs.(\ref{5}) in the form of a traveling wave assuming
that the junctions are long enough and neglecting reflected waves

\begin{equation}
\varphi_{1,2} = { \cal H}x - Vt + \psi_{1,2} + \theta_{1,2} + \theta_{1,2}^{(0)}
\label{7}
\end{equation}

Where $\psi_{1,2}$ is the rapidly oscillating part of $\varphi_{1,2}$ in space
with a period of 2$\pi/ \cal H $. $\theta_{1,2}$ is the slowly varying part and
$\theta_{1,2}^{(0)}$ = 0 (for 1) and $\pi$ (for 2). A similar representation was
used in \cite{r14} - \cite{r16} for finding the shape of the supersolitons
(dislocations) in the FL and in \cite{r8} for finding the spectrum of
acoustic-like oscillations of the FL. For the rapidly oscillating part
$\psi_{\pm} = (\psi_1 \pm \psi_2)$/2, we have from eqs.(\ref{5})

\begin{eqnarray}
\psi_+ = - \: \frac{sin(\theta_-)}{|D_+|^2} \left\{ b_+ cos(Y + \theta_+)  
+\alpha V sin(Y + \theta_+)\right\} \nonumber \\
\psi_- = \frac{cos(\theta_-)}{|D_-|^2} \left\{ - b_- sin(Y + \theta_+) +
\alpha V cos(Y + \theta_+)\right\} \nonumber \\
\label{8}
\end{eqnarray}

where Y = ${\cal H}x$ - Vt and $\theta_\pm$ = $(\theta_1 \pm \theta_2)$/2,
$b_\pm$ = $V^2_\pm$ - $V^2$, $D_\pm$ = $b_\pm +i\alpha$V and $V_\pm$ = $l_\pm
\cal H$. Assuming that $\psi_\pm$ are small (i.e.$|\psi_\pm|\ll1$), and
expanding sin($\varphi_\pm$) in powers of $\psi_\pm$, we obtain from
eq.(\ref{5}) the equations for the slowly varying part $\theta_\pm$ in the main
approximation 

\begin{eqnarray}
\l^2_+\partial^2_{xx}\theta_+  = (\partial^2_{tt} + \alpha\partial_t)\theta_+ 
+\alpha V A_- cos(2\theta_-)\nonumber \\ 
+\alpha V A_+ + \alpha V - \eta_+\nonumber \\
\l^2_-\partial^2_{xx}\theta_-  =  (\partial^2_{tt} + \alpha\partial_t)\theta_-
+ B sin(2\theta_-) - \eta_-
\label{9}
\end{eqnarray}

Here 4$A_\pm$ = $|D_-|^{-2} \pm |D_+|^{-2}$, 4B = $b_-|D_-|^{-2} - b_+|D_+|^{-2}$.
Eqs.(\ref{9}) describe the dynamics of the FL. The phase  $\theta_+$ is a 
local displacement of the FL as a whole and the phase $\theta_-$ determines
a relative  displacement of two fluxon chains.

{\bf II a. Collective modes}

Consider the stationary case when the currents $\eta_\pm$ are absent and the
FL is motionless (V = 0). Linearizing eqs.(\ref{9}), we obtain for the
spectrum of the collective modes (for simplicity we neglect the damping)

\begin{equation}
\omega^2 = \kappa^2 l^2_+\;,\; \omega^2 = \omega^2_0 + \kappa^2 l^2_-
\label{10}
\end{equation}

Here $\omega^2_0 = \gamma/[(1 - \gamma^2) l^2_J {\cal H}^2]$ is the threshold
frequency for the optical branch. It decreases with increasing $\cal H$. The
first expression in eq.(\ref{10}) describes the acoustic branch of the FL
oscillations. Similar modes exist in layered superconductors \cite{r6} -
\cite{r8}. These modes are independent from each other. In the acoustic
(optical) mode the phase $\theta_+$ ($\theta_-$) are perturbed. Consider now
excitations of these modes by the moving FL.

{\bf II b. Moving fluxon lattice}

Let us assume that a current $\eta_1 = \eta_2$ flows through both junctions
(then $\eta_+ = \eta_1 = \eta$ and $\eta_- = 0$). Then the stable solution of
eq.(\ref{9}) is: $2\theta_- = 0$ for $V < V_a, \;\;\; V > V_b \; $ and $2
\theta_- = \; \pi \; $ for $ V_a < V < V_b$. Here $V_{a,b}$ are the roots of
the equation $B(V) = 0$. The solution $\theta_- = 0, (\pi)$ corresponds to a
positive, (negative) value of B. For the case of a small damping $\alpha$ (i.e.
$\alpha \ll V_\pm$) we have $V^2_{a,b} \cong V^2_\pm\; [1 \mp \alpha^2/(V^2_+ -
V^2_-)]$

\begin{figure}
{\psfig{figure=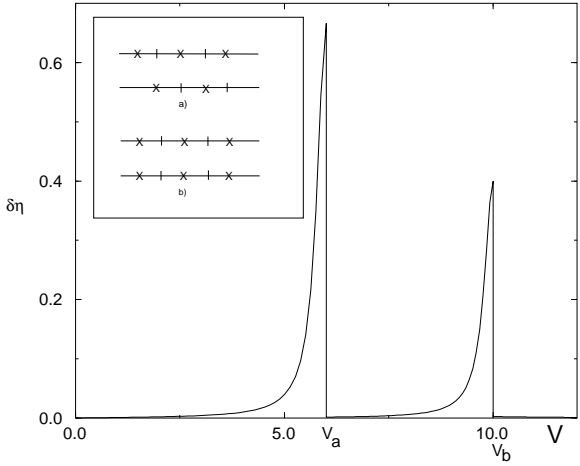,width=8cm}}
\caption{ The deviation of the current from Ohm's Law ($\delta\eta = \eta 
- \alpha$V) due to excitation of the collective modes vs voltage.(We used
$\alpha$ = 0.5) The positions of fluxons (crosses) in both junctions are
shown inset for $V > V_a$ ,$V_b < V$ (a) and $V_a < V < V_b$ (b).}
\label{f2}
\end{figure}

If the velocity of the moving FL exceeds ($V_a/\cal H$), a reconstruction  of
the moving FL occurs (see fig.\ref{f2}). If the velocity of the FL increases further
and exceeds ($V_b/\cal H$), the initial triangular form of the FL is restored.

The form of the current-voltage characteristics $\eta$(V) may be easily found 
from eq.(\ref{9}) averaged in space. We obtain

\begin{equation}
\eta - \alpha V = 2\alpha V \left\{ \begin{array} {r@{\quad,\quad}l}
|D_-|^{-2}  &  V < V_a \; ,\; V > V_b \\ |D_+|^{-2} & V_a < V < V_b
\end{array} \right.
\label{11}
\end{equation}

There are two peaks in the $\eta$(V) characteristics, one of them corresponds to
the velocity of the optical mode (at large $\kappa$) and the second corresponds
to the velocity of the acoustic mode. In addition there are two jumps at voltage
$V = V_a$ and $V = V_b$ corresponding to the FL reconstruction. If the damping
constant $\alpha$ is small, then $V_{a,b} \approx V_\pm$. In fig.\ref{f2} we present
the form of the $\eta(V)$ curve assuming for simplicity that $\alpha$ does not
depend on V.

{\bf II c. Creation of dislocations}

Let us assume that a bias current flows in the middle superconducting electrode
($\eta_- \neq 0$). One can see that the second equation of (\ref{9}) is similar
to an equation describing a single Josephson junction. This equation describes
nonlinear distortions in the FL. These distortions (dislocations) were analysed
in \cite{r14} where it was shown that they arise as kinks in the FL in two,
slightly different coupled Josephson junctions. Similar distortions
(supersolitons) may arise in a single long Josephson junction whose parameters
are modulated in space \cite{r15}. The characteristic length of dislocations
$l_-/B$ depends on both the magnetic field and the applied voltage (or current).
Dislocations are created at $\eta_- > B(V,\cal H$) and can propagate through the
system in an interval of $\eta_-$ above and below B (Fiske steps \cite{r9}). If
the damping is small, the dislocation has the well known fluxon form  $2\theta_-
= 4tg^{-1} exp (\xi)$, where $\xi$ = (x - ut)/$\l_d$ and $l_d$ = $l_-/\sqrt{B}$.
The velocity u is related to $\eta_-$ via the well known formula \cite{r16}
which in the slow velocity limit is reduced to u = $(\pi/4)\eta_-l_d/\alpha$.
According to eq.(\ref{9}), a dislocation arising in a moving FL causes a
perturbation of the phase $\theta_+$. Substituting the expression for $\theta_-$
into (\ref{9}), we obtain for $\theta_+$

\begin{equation}
\partial^2_{\xi\xi} \theta_+ + \beta\partial_\xi \theta_+ = -\; s\;\cdot
 cosh^{-2}\xi
\label{12}
\end{equation}

Where $\beta$ = $\alpha ul_d/[l^2_+ - u^2]$ and s = 2$\alpha V A_-\;\cdot
l^2_d/[l^2_+ - u^2]$. If the FL velocity V lies in the intervals (0,$V_a$),
($V_b,\infty$), we obtain from eq.(\ref{9}) $\eta_+ = \alpha V (A_+ + A_-)$.

From eq.(\ref{12}) we obtain

\begin{equation}
\partial_\xi \theta_+ = - \; s\int^{\xi}_{-\infty} d\xi_1 cosh^{-2} \xi_1\cdot
e^{\beta(\xi_1 - \xi)}
\label{13}
\end{equation}

\begin{figure}
{\psfig{figure=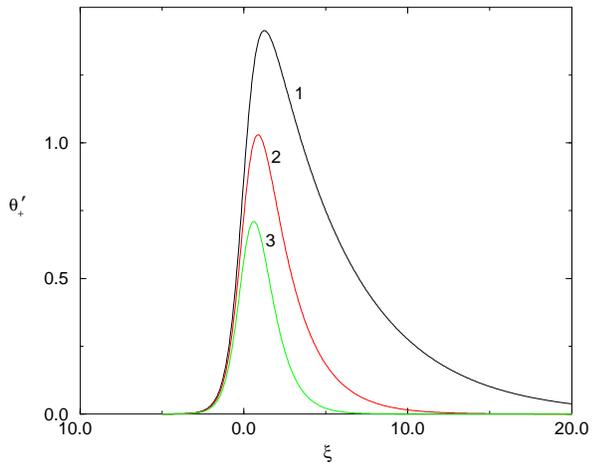,width=8cm}}
\caption{ The spatial dependence of the magnetic field ($\theta_+^\prime$ 
= $\partial_\xi \theta_+$) in the dislocation for different values of $\beta$:
$\beta$ = 0.2 (1), 0.2 (2), 1.0 (3).}
\label{f3}
\end{figure}

The spatial dependence of the dislocation $\theta_+(\xi)$ is shown in
fig.\ref{f3} for different values of $\beta$. The expression for $\theta_+(\xi)$
(\ref{13}) is valid provided that the characteristic size of the dislocation
$\beta^{-1}$ is less than the junction length L. Let us calculate the magnetic
flux carried by a dislocation in a moving FL. Substituting expression(\ref{13})
into eq.(\ref{3}), we have for magnetic flux in the system

\begin{equation}
\Phi = 2({\cal H} L) +2 \int dx\; \partial_x\theta_+
\label{14}
\end{equation}

The first term in eq.(\ref{14}) is the magnetic flux in the system in the 
absence of a dislocation. The second term is the magnetic flux $\Phi_d$
carried by a dislocation. With the help of eq.(\ref{13}) we obtain for $\Phi_d$

\begin{equation}
\Phi_d = -\; \frac{8VA_-}{u(V)} \; l_d
\label{15}
\end{equation}

One can see that the flux $\Phi_d$ depends both on the velocity of the
dislocation and the velocity of the FL as a whole, turning to zero at V = 0.
Therefore, a dislocation in the FL is a localized distortion which can move
under the action of an external force (the difference of the currents $\eta_-$)
and carry an arbitrary magnetic flux, the magnitude of which is determined by
currents $\eta_+$ and $\eta_-$.

{\bf III Conclusions}

We analysed the dynamics of a dense FL in a system of two coupled Josephson
junctions. Acoustic and optical collective modes may propagate in the FL. If the
FL is moving in the presence of a dc current, a resonance excitation of the
modes takes place when the FL velocity coincides with the limiting velocity
$V_-$ of the optical mode, or with the velocity of the acoustic mode $V_+$. In
the interval ($V_-,V_+$) the FL is  reconstructed. It was also shown that if the
currents $\eta_1$ and $\eta_2$ through the junctions are different, localized
distortions (dislocations) may be created in the FL. They carry an arbitrary
magnetic flux and may lead to a non-Josephson generation.

{\bf IV Acknowledgements}

We are grateful to the Royal Society, the Russian Fund for Fundamental Research
(project 96-02-16663a), the CRDF (project RP1-165) and also the EPSRC for  their
financial support.
One of the authors, (A.F.V.) wishes to thank J.Mygind  for the hospitality
shown to him during his visit to DTU (Lyngby, Denmark) where this work began.
We are also grateful to C.J. Lambert for the interest shown in this work.

\begin{figure}
{\psfig{figure=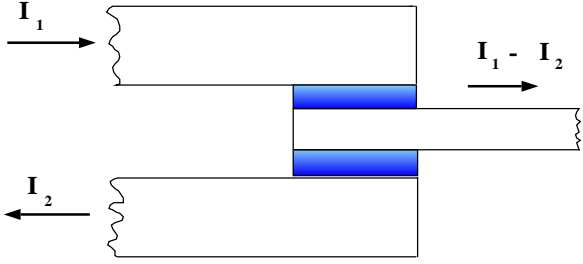,width=8cm}}
\caption{ The structure under consideration (schematically).}
\label{f1}
\end{figure}

\end{document}